\documentclass[a4paper,11pt]{article}
\usepackage{jinstpub} 
\usepackage{orcidlink}
\usepackage{xspace}
\usepackage{acro}
\usepackage{listings}
\usepackage{seqsplit}
\usepackage{cleveref} 
\crefname{figure}{figure}{figures}
\Crefname{figure}{Figure}{Figures}

\definecolor{lstkeyword}{rgb}{0.00,0.00,0.65}
\definecolor{lstcomment}{rgb}{0.40,0.40,0.40}
\definecolor{lststring}{rgb}{0.55,0.00,0.00}

\lstdefinestyle{python}{
    language=Python,
    basicstyle=\ttfamily\footnotesize,
    keywordstyle=\color{lstkeyword}\bfseries,
    commentstyle=\color{lstcomment}\itshape,
    stringstyle=\color{lststring}\ttfamily,
    showstringspaces=false,
    breaklines=true,
    frame=single,
    columns=fullflexible,
    xleftmargin=1em,
    xrightmargin=1em,
    numbers=left,
    numberstyle=\tiny\color{lstcomment},
    numbersep=8pt,
    stepnumber=1,
}

\lstdefinelanguage{g4macro}{
    morecomment=[l]{\#},
    commentstyle=\itshape\color{lstcomment},
}
\lstdefinestyle{g4macro}{
    language=g4macro,
    basicstyle=\ttfamily\footnotesize,
    commentstyle=\itshape\color{lstcomment},
    showstringspaces=false,
    breaklines=true,
    frame=single,
    columns=fullflexible,
    xleftmargin=1em,
    xrightmargin=1em,
    numbers=left,
    numberstyle=\tiny\color{lstcomment},
    numbersep=8pt,
    stepnumber=1,
}

\lstdefinelanguage{xml}
{
  basicstyle=\ttfamily\footnotesize,
  morestring=[b]",
  morestring=[b]',
  moredelim=[s][\bfseries\color{lstkeyword}]{<}{\ },
  moredelim=[s][\bfseries\color{lstkeyword}]{</}{>},
  moredelim=[l][\bfseries\color{lstkeyword}]{/>},
  moredelim=[l][\bfseries\color{lstkeyword}]{>},
  morecomment=[s]{<?}{?>},
  morecomment=[s]{<!--}{-->},
  commentstyle=\color{lstcomment},
  stringstyle=\color{lststring},
  showstringspaces=false
}

\DeclareAcronym{AI}{short = AI, long = artificial intelligence}
\DeclareAcronym{DOI}{short = DOI, long = digital object identifier}
\DeclareAcronym{GDML}{short = GDML, long = Geometry Description Markup Language}
\DeclareAcronym{HDF5}{short = HDF5, long = Hierarchical Data Format version 5}
\DeclareAcronym{HPGe}{short = HPGe, long = high-purity germanium}
\DeclareAcronym{PMT}{short = PMT, long = photomultiplier tube}
\DeclareAcronym{SiPM}{short = SiPM, long = silicon photomultiplier}
\DeclareAcronym{TCM}{short = TCM, long = time-coincidence map}
\DeclareAcronym{WLS}{short = WLS, long = wavelength-shifting}
\DeclareAcronym{CI/CD}{short = CI/CD, long = continuous integration and deployment}

\newcommand{\pkgformat}[1]{\texttt{#1}}
\newcommand{\remage}{\pkgformat{remage}\xspace}
\newcommand{\geant}{Geant4\xspace}
\newcommand{\mage}{\textsc{MaGe}\xspace}
\makeatletter
\def\hypenttloop#1 #2\hypenttend{%
  \seqsplit{#1}%
  \def\hypenttrest{#2}%
  \ifx\hypenttrest\@empty\else\ \hypenttloop#2\hypenttend\fi
}
\newcommand{\hypentt}[1]{\texttt{\hypenttloop#1 \hypenttend}}
\makeatother
\title{\texorpdfstring{\remage}{remage}: a Geant4-based simulation framework for low-background experiments}

\author[a]{T. Dixon~\orcidlink{0000-0001-8787-6336},}
\author[b]{E. Esch~\orcidlink{0009-0000-4920-9313},}
\author[d,e]{R. Henning~\orcidlink{0000-0001-8651-2960},}
\author[c]{M. Huber~\orcidlink{0009-0000-5212-2999},}
\author[c]{M. Neuberger~\orcidlink{0009-0001-8471-9076},}
\author[c,1]{L. Pertoldi~\orcidlink{0000-0002-0467-2571}%
\note{Corresponding author.}}
\author[d]{and Z. Wang~\orcidlink{0009-0007-4386-0819}}

\affiliation[a]{Department of Physics and Astronomy, University College London, London WC1E 6BT, United Kingdom}
\affiliation[b]{Department of Physics, University Tübingen, Tübingen 72076, Germany}
\affiliation[c]{Department of Physics, TUM School of Natural Sciences, Technical University of Munich, 85748 Garching b.~München, Germany}
\affiliation[d]{Department of Physics and Astronomy, University of North Carolina, Chapel Hill, North Carolina 27599, USA}
\affiliation[e]{Triangle Universities Nuclear Laboratory, Durham, NC 27708, USA}

\emailAdd{luigi.pertoldi@tum.de}

\abstract{%
Monte Carlo simulations are an essential tool for data modeling and experimental design in low-background physics. These simulations typically demand a considerable effort, with each collaboration developing and maintaining its own complex \geant application. In this paper we present the design and validation of \remage, which is an open-source framework designed to standardize these simulations and significantly reduce this effort. Its design separates the geometry, built with dedicated Python tools, from the compiled application. It provides a physics list and event generators tuned to radiogenic and cosmogenic backgrounds, output tailored to the most common detector types, and robust validation backed by modern software development practices. The software is distributed ready to use and scales efficiently with the computing resources that demanding simulations require. \remage has been primarily developed for the LEGEND experiment, but is applicable to a wide variety of experiments.%
}


\arxivnumber{1234.56789} 

\begin{document}
\maketitle
\flushbottom

\section{Introduction}%
\label{sec:intro}

Monte Carlo simulations are a crucial tool in high-energy physics, both for predicting experimental outcomes and for planning new experiments.
\geant is a state-of-the-art particle tracking toolkit for nuclear, leptonic and optical simulations~\cite{GEANT4:2002zbu,Allison:2006ve,Allison:2016lfl}, which are ubiquitous in experimental particle physics and other fields such as medical physics. For example, they are used in low-background physics experiments to simulate radioactive and cosmogenic (i.e., induced by cosmic radiation) background sources, for experimental design and data analysis (see for example~\cite{GERDA:2019cav,CUORE:2024fak,LZ:2022ysc,GERDA:2022hxs,Majorana:2024ehx,Wiesinger:2018qxt,Pec:2023yic}).

\geant is a general toolkit supporting a vast range of uses. For a particular use case it is typically necessary to develop a dedicated \geant application. A significant amount of effort goes into the development of these programs. This includes the implementation of specific (often complex) experimental geometries, selection of appropriate physics processes, and implementation of routines to generate primary particles and store output information. Since \geant is based on C++, a low-level programming language with a large barrier to entry, this typically implies writing and maintaining large amounts of C++ code, which further increases the effort required. Finally, considerable validation is required to confirm that the application is appropriate for a certain use case, in particular the physics processes selected from such a general toolkit. The development of these \geant applications has historically been performed separately by each experiment or research group~\cite{Boswell:2010mr,CUORE:2024fak,She:2021dag,Angloher:2026wfr}, with sharing limited by the presence of sensitive information in these applications.

\remage is a \geant-based framework designed to reduce the effort required to run Monte Carlo simulations for low-background physics experiments. Unlike most \geant applications of this kind, its design separates the definition of the experimental geometry from particle generation and tracking. \remage is inspired by the \mage~\cite{Boswell:2010mr} project, which enabled sharing of simulation routines between the \textsc{Majorana Demonstrator} and \textsc{Gerda} experiments. We generalize this approach to allow use by a wide range of low-background physics experiments, simplify the definition of experimental geometries, and employ modern software development practices. This approach also standardizes the selection of physics processes appropriate for low-background experiments, based on a dedicated sequence of validation routines that are fully integrated into the software design and run automatically.
This specialization for low-background physics also covers the computational performance necessary to investigate rare background sources and flexible output structures.
Other projects either target different applications, such as beamline simulations~\cite{Nevay:2018zhp}, or do not provide enough functionality for low-background physics experiments~\cite{Arce:2014loa,Liu:2025ujy,g4simple}.

This paper describes the design and implementation of the \remage framework. In \cref{sec:design}, we introduce the design principles of the framework. \Cref{sec:geometry,sec:generators,sec:physicslist,sec:running,sec:output} describe its main components: defining experimental geometries, generating initial particles, selecting appropriate physics processes, running the simulation, and storing and reducing the outputs. \Cref{sec:validation,sec:development,sec:usage} then cover the end-to-end validation of \remage, the software development and the applications of the framework. Throughout, we describe \remage v1.2.0~\cite{remage_v12} used with \geant version 11.3.2, the most recent \geant version our validation suite had been run against at the time of writing.

\section{Design principles of the \texorpdfstring{\remage}{remage} framework}%
\label{sec:design}

\begin{figure}
    \centering
    \includegraphics[width=\textwidth]{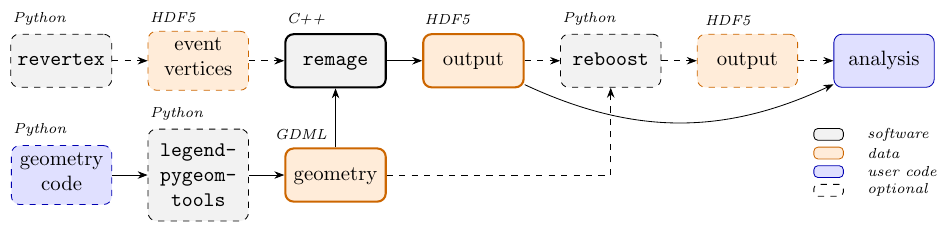}
    \caption{%
    The main \remage simulation flow, with the typical processing steps and the interfaces between them. The core of the workflow is drawn with solid arrows: the \remage application reads the experimental geometry as a \acs*{GDML} file and writes the simulation results in \acs*{HDF5} format. Everything drawn with dashed lines is optional and lives outside the core application in the surrounding ecosystem of packages that produce inputs (\pkgformat{revertex}) or post-process the output (\pkgformat{reboost}). Since the interfaces are well-defined file formats, these components can be replaced by any other tool utilizing the same data exchange format.}\label{fig:remage-flow}
\end{figure}

\remage is built on a design and structure that differs from most typical \geant applications. In this section we describe it, leaving the details of the individual components to the following sections.

The guiding principle of this design is to limit the compiled application to the tasks that strictly depend on \geant. We move everything else, for example geometry definition, into separate packages, which can be written in programming languages other than C++, such as Python. This modular approach minimizes the amount of C++ code that has to be written and maintained. \remage is therefore both this core C++ application and, by extension, the whole framework it forms together with these packages. This modular approach means that \remage does not contain any geometry code, which usually carries sensitive experiment-specific information, and can therefore be released as open-source software. The separation also allows the packages to be maintained independently of the main application. The general design of the resulting software, with its modular components and data exchange formats, is shown in \cref{fig:remage-flow}.

 Building \remage on wrappers that expose \geant from a higher-level language, such as the Python bindings~\cite{Geant4Py} or Geant4.jl~\cite{Geant4jl,Stewart:2025pwu} for Julia, would have been a legitimate alternative, but both target simple applications and are not yet mature enough for an advanced application like \remage.

Separating the C++ application from the surrounding packages depends on a clear interface between the individual components. We therefore define that interface in terms of the data exchanged across it, rather than in terms of code. The experimental geometry enters the application as a \ac{GDML}~\cite{Chytracek:2006be} file, and the simulation results leave it in one of several supported output formats. Among these, \remage recommends its own \ac{HDF5}~\cite{HDFGroup:2025} layout, following the LEGEND \ac{HDF5} specification~\cite{legend-data-format-specs_2020}, as the primary output format, since only this format exposes the framework's most advanced output features, described in \cref{sec:output}. Optionally, input particle kinematics and positions (``event vertices'') can also be read by \remage as \ac{HDF5} files (\cref{sec:vertex,sec:generators}), which enables event generators to be written in languages other than C++. We stay as close to the plain formats as we can: \remage reads standard \ac{GDML}, and the few additional constructs it relies on are documented precisely (\cref{sec:geometry}), as are the data structures it writes (\cref{sec:output}). These formats are language-agnostic, so every step before and after the tracking simulation can be implemented in whichever language suits it best, in our case usually Python or Julia.

Within this boundary, the executable has to be flexible enough that users do not need to modify or rebuild it. Most of \remage's features are therefore exposed through a command line interface (\cref{sec:running}) and through a rich set of \geant messenger commands, which can be collected in macro files; \cref{lst:remage-macro-example} in \cref{app:example} shows such a macro file for a typical simulation setup. \remage's own commands are explicitly designed to keep these files simple, which reduces both the work of writing them and the risk of mistakes. Evaluating a different geometry version, physics scenario or set of output options becomes a matter of configuration rather than recompilation, which keeps the turnaround short and avoids duplicating large amounts of code when building upon an existing simulation setup. In particular, to support a broad range of low-background physics applications, the output format is strongly customizable, as described in \cref{sec:output}.

Since most users interact with \remage via \geant macro commands, we place a particular focus on documentation as a critical component of the software design; one important part of this is the automatic generation of documentation for all \remage macro commands~\cite{remage-docs}. Needs that go beyond the available commands are still accommodated, since the application logic is also available as a library that users can extend with their own C++ code, as described in \cref{sec:development}.

Users who do not have to modify the application should not have to compile it either, and we treat this as a deliberate design goal rather than an afterthought. We provide container images for the \remage core and conda-forge~\cite{conda-forge} packages, so that even the compiled C++ application is available ready to use, and a full simulation can be installed and run without ever needing to set up a C++ build environment (\cref{sec:development}). The data exchange layer is in turn only as useful as the tools available to work with it. We therefore also maintain the packages that surround it: \pkgformat{legend-pygeom-tools}~\cite{legend-pygeom-tools_2026}, built on \pkgformat{pyg4ometry}~\cite{Boogert:2020cmn,pyg4ometry_2025}, for writing geometries as Python code, \pkgformat{revertex}~\cite{revertex_2025} for generating input vertices (\cref{sec:vertex}), \pkgformat{legend-lh5io}~\cite{legend-lh5io_2026} for reading and writing the primary \ac{HDF5} output, and \pkgformat{reboost}~\cite{reboost_2025} for post-processing the output (\cref{sec:postproc}). 

Validation of the simulation outputs and methods is essential for enabling robust and reliable results, and again we treat this as a critical component of the software design, backed by a dedicated set of software tests and validation routines. This encompasses tests of the individual algorithms, checks on physics processes (\cref{sec:physicslist}) and full end-to-end validation by comparing to experimental data (\cref{sec:validation}).
We use GitHub continuous integration to run these tests and simulations automatically for every change to the \remage software, and the results are then stored on a website as part of the \remage documentation~\cite{remage-validation}. This means that the validation is naturally updated for every change in \remage or \geant, which is what makes it possible to adopt each new \geant version while keeping the physics output stable. It has already allowed us to uncover regressions in the upstream \geant project, which we reported to its developers~\cite{Geant4Bugzilla2759,Geant4Bugzilla2750}.

Low-background physics searches often require simulations with high statistics in order to study rare background sources. \remage's performance is therefore an important consideration for use on both personal computers and high-performance computing systems. In addition to the built-in \geant multi-threading, \remage implements a custom multi-process parallelization mode to achieve consistent performance scaling on large machines; both parallelization modes are described in \cref{sec:running}. Runtime and output volume can be reduced further through step limits (\cref{sec:physicslist}), the clustering of \geant steps (\cref{sec:clustering}), and the choice of output fields (\cref{sec:outputschemes}).

\section{Experimental geometry}%
\label{sec:geometry}

In \geant the detector construction is defined as a hierarchy of nested volumes, together with the materials they are made of and their physical properties. For a non-trivial experimental setup, this geometry definition is typically the largest part, by volume of code, of a \geant application, and also the part that changes most frequently as a detector design evolves.
In a typical \geant setup, every such change requires a rebuild of the application.

Alternatively, the geometry can be described in a dedicated text-based input format, such as \geant's own ASCII geometry description, used for example by GAMOS~\cite{Arce:2014loa}. This removes the need to recompile, but it asks the user to learn a dedicated syntax, and such files are typically written by hand.

As mentioned previously, \remage avoids this by not implementing the geometry as C++ code at all: the volume tree is instead loaded at run time from a user-supplied \ac{GDML} file, a widely used, application-independent XML-based file format that closely follows the geometry object model implemented by \geant~\cite{Chytracek:2006be,GDMLUserGuide}, and for which several other reading and writing software toolkits in different programming languages exist~\cite{Brun:1997pa,Boogert:2020cmn,pyg4ometry_2025,FreeCADGDML}. The format can describe the experimental geometry and materials. In addition, it can specify the optical properties of the setup, meaning optical photons can be simulated (see \cref{sec:physicslist}) without additional inputs. \ac{GDML} files are typically written by a user's geometry-writing application, and not by hand.
\ac{GDML} files, however, do not typically carry any information on which volumes are sensitive detectors, on how volumes should be drawn, or on whether and how \geant regions are defined. In the following we describe the custom extension that carries this information alongside the geometry (\cref{sec:detectors}), how both are most conveniently produced together in Python (\cref{sec:pygeom}), and how the result can be checked and profiled before it is used (\cref{sec:geomcheck}).

\subsection{Custom \texorpdfstring{\acs{GDML}}{GDML} extensions}%
\label{sec:detectors}

Closely related to the geometry definition is the assignment of detectors to volumes, i.e., to make them respond in a specific way to particles interacting inside them, and to store this information into output files. Throughout this paper we use ``detector'' for what \geant itself calls a sensitive volume, since this is the term \remage itself uses. This assignment is mostly relevant for reading out the hits recorded in these volumes, and a complete description of how it shapes the simulation output is given in \cref{sec:output}; here we focus on how detectors are defined alongside the geometry.

Detectors can be registered at run time through a macro command, which assigns a detector type and a user-defined unique identifier to a volume, used internally by \remage and to label the detector in the output. For complex setups, which usually reuse the same large set of detectors across most of their simulation workflows, it is more convenient to store this assignment in the geometry file and to create it together with the geometry itself. Plain \ac{GDML} cannot carry it, however, since detector logic is not part of the \geant geometry model that \ac{GDML} describes.

\remage closes this gap with a custom extension that does not modify the \ac{GDML} schema itself, but reuses \geant's own standard mechanism for including arbitrary auxiliary values in a \ac{GDML} file~\cite{GDMLUserGuide}. Files produced with this extension therefore remain valid, plain \ac{GDML}, readable by any \ac{GDML}-capable viewer or application, while carrying additional structures that \remage understands. Detector assignment uses the \texttt{RMG\_detector} and \hypentt{RMG\_detector\_meta} auxiliary structures, which associate a physical volume with a detector type, a numeric unique identifier, and further optional metadata, e.g.,

\begin{lstlisting}[language=xml, frame=none, numbers=none]
<userinfo>
  <auxiliary auxtype="RMG_detector" auxvalue="germanium">
    <auxiliary auxtype="det_pv" auxvalue="101"/>
  </auxiliary>
  <auxiliary auxtype="RMG_detector_meta" auxvalue="">
    <auxiliary auxtype="det_pv" auxvalue='{"manufacturer": "D. Radford"}'/>
  </auxiliary>
</userinfo>
\end{lstlisting}

defines the physical volume \texttt{det\_pv} as a \texttt{germanium}-type detector with unique identifier 101, just as registering it at run time would. The \hypentt{RMG\_detector\_meta} entry attaches optional, user-defined metadata, which can carry detector information that has no equivalent in \geant's own geometry model and can be read from the \acs{GDML} file by other tools.

This detector setup is not read from the \ac{GDML} file automatically, but only on request through a dedicated macro command. Registration at run time with macro commands remains useful for volumes not authored that way, for interoperability with \ac{GDML} files that lack this metadata, or for registering and overriding detectors at run time without touching the geometry file.

A second structure, \texttt{rmg\_color}, stores the RGBA color and transparency of a logical volume, read only by our visualization tools (\cref{sec:pygeom}).
A third structure, \texttt{Region}, encodes a grouping of logical volumes into a \geant region~\cite{Geant4:Regions}, together with associated production cuts and step, track, time or energy limits, in the format expected by \geant's own region-import mechanism; this structure can be written directly, and \remage activates it automatically at run time when the geometry is loaded.

\subsection{Practical geometry construction in Python}%
\label{sec:pygeom}

In practice, both the geometry and its auxiliary metadata are most conveniently produced together in Python. To produce these \ac{GDML} files, we rely on \pkgformat{pyg4ometry}~\cite{Boogert:2020cmn,pyg4ometry_2025}, a general-purpose, \geant-independent Python geometry package. Its Python classes for constructive solids and logical and physical volumes closely mirror \geant's own C++ interface, so that users familiar with it can easily adapt their work; and it supports reading, writing and editing \ac{GDML} files directly. Most features of the geometry model are supported: solids, volume structure, materials and optical material properties. \pkgformat{pyg4ometry} can also import CAD geometries, with tessellation, and ROOT geometries, converting both to \ac{GDML} and making it straightforward to reuse existing engineering or detector models. 
It further includes a standalone, VTK-based~\cite{Schroeder:2006} interactive visualization that runs without any \geant installation, useful for users who only want to inspect a geometry rather than simulate it.

We extend \pkgformat{pyg4ometry} with our own package, \pkgformat{legend-pygeom-tools}~\cite{legend-pygeom-tools_2026}, for the \remage-specific parts of this workflow. Besides the detector-registration functionality described below, it provides a single function that commits all pending auxiliary metadata, detectors, colors and regions, and writes the final \ac{GDML} file. Before writing, a set of geometry sanity checks is run, covering registry consistency and common material definition mistakes, such as not registering all components with the \pkgformat{pyg4ometry} registry before writing the file. For visualization, it provides the standalone command-line viewer \texttt{legend-pygeom-vis}, which extends \pkgformat{pyg4ometry}'s VTK-based display with a configurable YAML ``scene file'' format, supporting camera presets, clipping planes, per-volume color overrides, overlay of points or events read from the \ac{HDF5} output, and offscreen export to PNG images.

On the Python side, it is possible to attach to a physical volume object the information \remage requires for detector registration. \Cref{lst:pygeom-example} in \cref{app:example} shows the detector registration in a more complete workflow, together with attaching a visualization color, defining a \geant region, and iterating over the registered detectors before writing the final \ac{GDML} file.

See \cref{sec:usage} for an outlook on the existing Python packages that implement, on top of this ecosystem, the geometry of specific experimental setups.

\subsection{Geometry checking and benchmarking}%
\label{sec:geomcheck}

\remage provides three distinct mechanisms to validate and profile a geometry, addressing different needs and offering different levels of detail. By default, every \ac{GDML} file loaded by \remage is checked for volume overlaps using \geant's own sampled-point overlap test. This always-on check is a convenient first line of defense against basic geometry construction errors.

A deeper, opt-in correctness check is available through the \hypentt{GeometryCheck} output scheme (see \cref{sec:outputschemes}). Unlike the overlap check, this mechanism is based on geantinos: massless, chargeless, non-interacting \geant tracking particles used purely to probe navigation. \remage fires geantino primaries through the geometry and verifies, at every step, that the pre- and post-step volume transitions reported by the navigator are consistent with the declared volume hierarchy. This catches subtler navigation bugs than a plain overlap test can, such as cases where volumes of the geometry are only overlapping through coincident surfaces, and not their bulk.

\begin{figure}
    \centering
    \includegraphics[width=\textwidth]{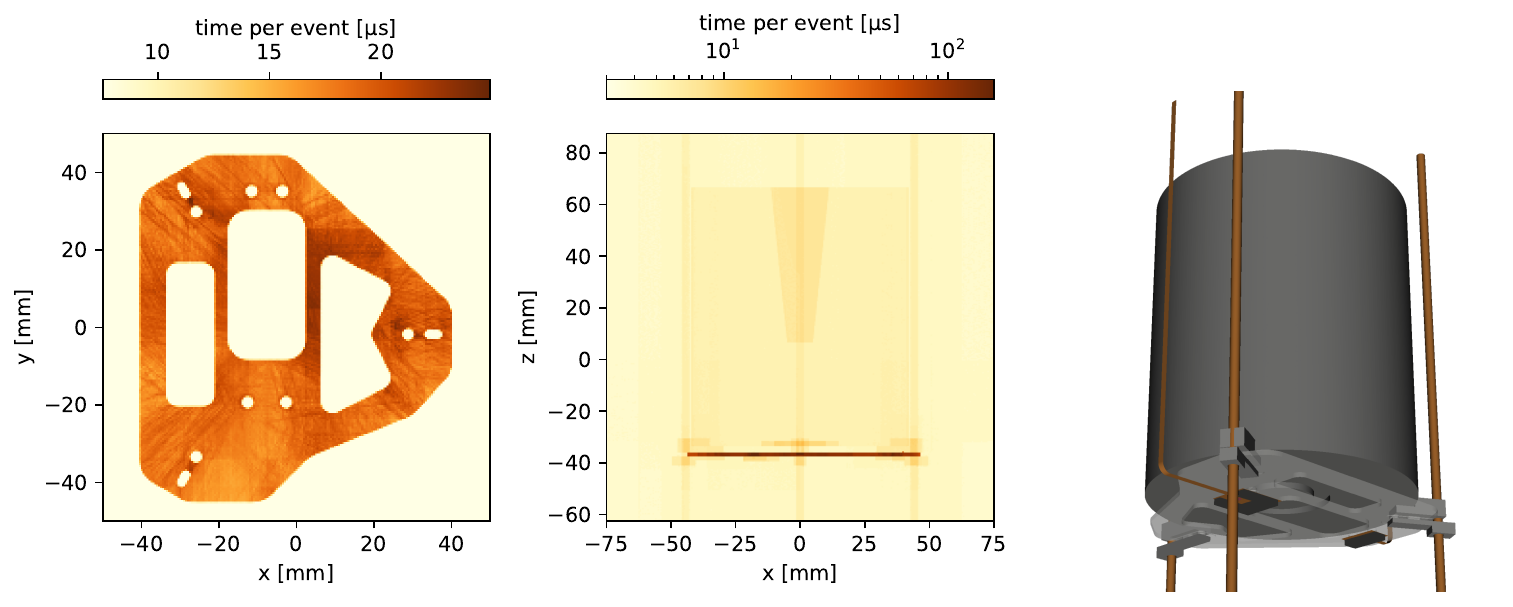}
    \caption{%
    Output of \texttt{remage-geombench} (left panels) for a sample geometry (rendered in the rightmost panel), a LEGEND-1000 detector unit~\cite{legend-pygeom-l1000_2026}. Each point of a map gives the wall-clock time \geant spends navigating a single geantino launched from that point through the geometry, for two orthogonal sampling planes. The first isolates the base plate below the germanium detector, a tessellated solid whose many facets make tracking slow; the second shows it alongside parts built from \geant native solids, where it still dominates the simulation time.
    \label{fig:geombench}}
\end{figure}

\remage also ships a standalone command-line tool, \texttt{remage-geombench}, for profiling the navigation performance of a geometry rather than its correctness, which is useful during geometry design to identify components that slow down navigation. Like the \hypentt{GeometryCheck} scheme, it relies on geantinos, but shoots them from a regular three-dimensional grid of starting points laid out on three orthogonal sampling planes, and records the navigation wall-clock time at each grid point. The resulting per-point timing map highlights geometry regions where navigation is disproportionately slow, guiding further optimization before running full physics simulations. An example output for a sample geometry with multiple different volumes is shown in \cref{fig:geombench}.

\section{Primary event generation}
\label{sec:generators}

With the geometry defined by the user, the next key aspect of a \geant simulation is to define the primary particles and their initial state. Every simulated event starts from a primary vertex: the position and time at which one or more particles are created, together with each particle's type and momentum. Producing all of these quantities is the task of what is generally referred to as an event generator.

In low-background experiments, the spatial distribution of primary vertices deserves particular attention of its own, since reproducing a radiogenic background correctly requires knowing not only what decays, but where, within the experimental setup, it occurs. \remage reflects this by conceptually splitting the definition of an event's initial state into two separate steps: spatial sampling, referred to as ``vertex generation'', and the generation of the primary particle kinematics, referred to as ``event generation''; these are configured independently of each other and are described separately in the following two subsections, with the caveat that some event generators produce the vertex position themselves. \geant ships with general-purpose primary generators, but these alone rarely cover the needs of a realistic experiment, in either of the two steps. \remage therefore extends them, integrating a number of generators tailored to the needs of low-background experiments, some of which are based on sophisticated algorithms. 

\subsection{Vertex generators}\label{sec:vertex}
Low-background experiments typically need to account for several sources of potential radioactive backgrounds originating from their experimental setup, which are routinely quantified with Monte Carlo-based background models~\cite{GERDA:2019cav,Majorana:2024ehx,CUORE:2024fak,LZ:2022ysc}. Depending on the source of the radioactive contaminant, it will manifest itself either in the bulk of a material or only in a thin layer at its surface, for instance as a result of handling or machining during detector fabrication. \remage provides dedicated vertex confinement modes that place primary vertices to reproduce where such contamination occurs, described in the following. By design, every simulated event has exactly one vertex, shared by all particles produced in that event. Note that some of the generators described in \cref{sec:kinematics} can also provide vertex positions on their own; in that case, no separate vertex confinement needs to be configured.

The volumes to sample from can be either physical volumes already present in the \geant geometry or virtual, independently defined solids. Physical volumes are selected by name or, more generally, by regular expression; virtual solids, such as boxes, spheres or cylinders, are defined by the user through \remage macro commands.
Since a contamination normally affects all components of the same type, and these are usually named after a common scheme in the geometry, a single pattern such as \hypentt{det\_holder\_.*} selects them all at once, keeping the macro file short and sparing the user the error-prone enumeration of hundreds of volume names.
For some simple constructive solids (box, sphere, orb and tube) \remage samples analytically; for any other \geant solid, it falls back to a generic rejection-sampling algorithm, using the solid's bounding extent together with an inside/outside containment check. Sampling from the bulk of these volumes is the default; sampling on their surface can be selected instead and is described below. 

\begin{figure}
    \centering
    \includegraphics[width=\textwidth]{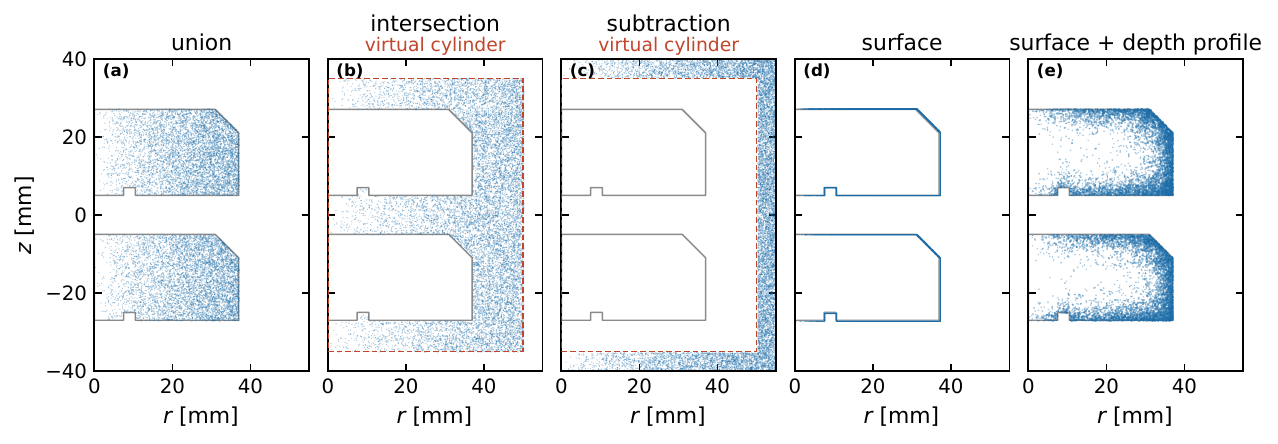}
    \caption{%
      Some vertex confinement options provided by \remage, illustrated with two germanium detectors in a medium. The figures show the radial component of the sampled vertices and thus exhibit a $1/r$ dependence in the density. From left to right: (a) shows bulk sampling in both detectors (union). (b) samples from the intersection of the surrounding volume and a virtual, user-defined cylinder. In (c), this cylinder is rather subtracted from the surrounding volume. (d) and (e) show the surface samples, both without and with an additional depth distribution applied to the sampled vertices.
    \label{fig:confinement}}
\end{figure}

When several volumes are configured in this way, the method used to combine them can also be chosen from three modes. The default mode samples from the union of all of them, and is the natural choice when several detector components share the same contamination scenario, as shown in~\cref{fig:confinement}a. The other two combination modes handle cases where the contamination is only present in a subregion of a volume: sampling can be restricted to the intersection of the physical volumes with a user-defined virtual solid (\cref{fig:confinement}b), or, conversely, one or more virtual solids can be excluded from the sampling region (\cref{fig:confinement}c).

By default, the vertex density is uniformly sampled across the sampling region regardless of how many volumes contribute to it. Two alternative weighting schemes are available for cases where the contaminant is not expected to scale with the geometric volume: the vertex density can instead be weighted by the volume mass, or by the mass contribution of one isotope. The latter is the natural choice when the simulated process is bound to the abundance of an isotope, for instance decays generated (\cref{sec:kinematics}) in a set of volumes with different, known enrichment fractions. These mass-weighting schemes apply only to bulk sampling; surface sampling is always weighted by surface area instead, and cannot be combined with them.

For some simple constructive solids (box, sphere, orb and tube) \remage also can sample vertices on the surface analytically.
Surface sampling of arbitrary solids instead relies on an algorithm based on counting the intersection points of lines with the surface, as described in~\cite{Detwiler:2008bs}. Unlike the description in that publication, \remage directly implements this surface sampling algorithm without using a first simulation pass. Surface sampling is shown in~\cref{fig:confinement}d.

Surface-sampled vertices can optionally be displaced inward into the material, to model sub-surface contamination such as implanted or diffused activity, as introduced by machining of parts or surface treatments. A depth profile can be configured, choosing between an exponential, a uniform or a truncated Gaussian shape, with associated parameters; the sampled depth is then subtracted from the sampled surface point along the inward normal, as shown in~\cref{fig:confinement}e. The sampled depth of the displaced vertex is clamped to the distance to the opposite surface of the volume to avoid placing vertices outside of the volume.

All the sampling algorithms described above are tested automatically by the \remage validation suite, which verifies the distribution of the generated vertices both statistically and through visual checks.

For scenarios not covered by these built-in options, \remage can instead read event vertices from an external file; the specification of the input format and further implementation details are given in the \remage documentation~\cite{remage-docs}. The Python package \pkgformat{revertex}~\cite{revertex_2025} generates compatible files, covering both vertex positions and, as described in \cref{sec:kinematics}, primary particle kinematics. One case the built-in sampling does not cover, but \pkgformat{revertex} does, is sampling of vertices on a specific surface of a volume only, e.g., on the differently processed surfaces of an \ac{HPGe} detector that are differently affected by some background components.

\subsection{Event generators}\label{sec:kinematics}
The simplest way to generate primary events in \remage is to reuse \geant's own built-in generators, the general particle source or the more basic particle gun. Beyond these, \remage directly includes the C++-based cosmogenic muon generator EcoMug for surface muon fluxes~\cite{Pagano:2021wqu}, and links against BxDecay0~\cite{Tretyak:2017}, which provides both double-$\beta$ decay and single-isotope background decay generators; both can be configured directly with macro commands. \remage also supports reading input files generated by MUSUN, which samples cosmogenic muons from the distributions specific to underground laboratories~\cite{Kudryavtsev:2008qh}. These specialized generators are crucial for simulating cosmogenic backgrounds in both above-ground and underground experiments, as well as for the physics requirements of experiments searching for double-$\beta$ decay.

No set of generators can cover every case, so \remage also accepts primary particle kinematics from an external file, produced by whichever tool or script suits the problem. This makes it possible, for example, to study exotic physics processes that no existing generator implements, without modifying the core \remage application. The format is the same \ac{HDF5}-based one already introduced for input vertices in \cref{sec:vertex}, with a table providing columns for the particle type as a PDG code~\cite{ParticleDataGroup:2024cfk}, its kinetic energy, momentum direction and emission time. As for vertex files, the full format specification and further implementation details are given in the \remage documentation~\cite{remage-docs}.

Vertex confinement and kinematics generation are two independent configuration choices in \remage, and can be combined freely: BxDecay0 kinematics, for instance, can be distributed over an array of detector volumes to simulate double-$\beta$ decay, or paired with vertices read from an external vertex-only file; conversely, a kinematics-only file can be combined with volume-based vertex confinement. A single, combined table carrying both position and kinematics columns together is also supported. Besides the vertex-generation routines described in \cref{sec:vertex}, \pkgformat{revertex}~\cite{revertex_2025} also provides dedicated primary-kinematics generators, writing their output directly in the format \remage reads back in.

\section{Physics list}%
\label{sec:physicslist}

Once the primary particles have been generated, the physical processes they can undergo, and the models used to simulate them, must be specified. In \geant, the set of enabled physics processes and their parameters are bundled in a ``physics list''~\cite{Geant4:PhysicsListGuide}. \remage provides a custom physics list tuned to the needs of low-background experiments, namely the simulation of radiogenic backgrounds in the energy range from keV to a few MeV, as well as cosmogenic backgrounds.
It covers the electromagnetic interactions of $\gamma$-rays and charged particles, radioactive decays and their chains, the hadronic processes induced by cosmic-ray muons and neutrons, and the production and propagation of scintillation and Cherenkov light. Its parameters are furthermore chosen to reach the precision required to model the response of solid-state and optical detectors. The list is close to the default physics list in upstream \geant, and is mostly built from the physics building blocks (``physics constructors'') that \geant itself provides, with some notable changes. In terms of physics content it also closely resembles the list used by \mage~\cite{Boswell:2010mr}, which addressed the same class of experiments, although \remage implements it independently and targets more recent \geant versions.

The \remage validation suite covers every part of this physics list, comparing important features of the simulated processes to experimental data, to theoretical calculations, or in some cases to other simulation frameworks. In the following we describe the physics list and, alongside it, show a selection of these checks.

\paragraph{Electromagnetic physics.}
By default, electromagnetic interactions are simulated with the ``Livermore'' constructor (\hypentt{G4EmLivermorePhysics})~\cite{Geant4:LivermoreConstructor,Geant4:LivermoreModel}, which is one of the \geant electromagnetic options that are specifically also targeted at low-energy applications. Its models are built on evaluated atomic data libraries, describe $\gamma$ and electron interactions down to 100~eV, and account for atomic effects such as fluorescence and Doppler broadening. This low-energy treatment concerns $\gamma$ rays and electrons: positrons, muons and ions are described as in \geant's standard electromagnetic models.
A macro command lets the user select one of the other electromagnetic constructors shipped with \geant instead of the Livermore constructor, among a fixed set of alternatives.
Like most upstream physics lists, \remage enables the simulation of electro-nuclear, muon-nuclear and gamma-nuclear interactions (\hypentt{G4EmExtraPhysics}).

\paragraph{Decay physics.}
Both the decays of unstable particles (\hypentt{G4DecayPhysics}) and the radioactive decays of nuclei (\hypentt{G4RadioactiveDecayPhysics}) are simulated by default. \geant does not simulate the decay of a nuclide whose lifetime, or whose sampled decay time, exceeds a configurable threshold, effectively treating it as stable, and \geant's recent default value of one year is far too short for the long-lived members of the natural decay series. \remage therefore raises the threshold to \geant's previous default value exceeding the age of the universe ($10^{18}\,\mathrm{s} \approx 3.17\times10^{10}\,\mathrm{yr}$), so that entire chains can be followed, and leaves it adjustable through a macro command. \remage additionally enables the angular correlations between the $\gamma$ rays emitted in a de-excitation cascade, which \geant leaves switched off by default. Decays that \geant's own data cannot describe, double-$\beta$ decay in particular, are instead handled by the dedicated generators discussed in \cref{sec:generators}, such as BxDecay0.

\remage also implements a custom, optional process for inner bremsstrahlung that \geant does not model. Inner bremsstrahlung is the radiation emitted by the sudden acceleration of the electron as it is created in a $\beta$ decay, rather than by its later scattering in matter~\cite{Hayen:2017pwg}. Conversely, inner bremsstrahlung radiation can also be emitted during an electron capture decay, which is not implemented in \remage. Currently this implementation is experimental and not enabled by default.

\paragraph{Production cuts and step limits.}
\geant will generally track all created particles until they lose all their energy. Some processes may create many secondary particles with low energies, and tracking these would lead to excessive simulation time and output file size. Setting production cuts remediates this, as low-energy particles are not produced for tracking if they have less than a specified range left to travel, as determined by their kinetic energy~\cite{Geant4:Cuts}. The energy is instead locally deposited. This stops a large number of low-energy particles (e.g., low-energy bremsstrahlung photons) from increasing simulation time without significantly changing the simulation result.

By default, the \remage physics list uses a production cut of 0.1~mm for all particles, which is lower than the cuts in all upstream physics lists and corresponds to a production threshold of about 175~keV for electrons and 6~keV for gammas in germanium. This is necessary for high-precision modeling of signal generation in germanium detectors, as these are sensitive to the position of the ionizing interactions on this scale. This can be altered via \remage macro commands or, for individual groups of volumes, through the region definitions carried by the geometry file itself (\cref{sec:detectors}).

\remage further allows a maximum step length to be imposed on the tracking of charged particles separately for each volume through a macro command. No step limit is set by default, in which case \geant derives the step length from the distance to the next volume boundary, the mean free path to the next interaction, and the multiple-scattering algorithm~\cite{Ivanchenko:2010zz}. Forcing shorter steps samples the energy deposition along a track more finely, at a cost in computing time and output volume that grows steeply.
This matters wherever the spatial distribution of the deposits is important, such as in the thin region at the surface of an \ac{HPGe} detector where the charge collection efficiency is sensitive to the precise location.

\begin{figure}
    \centering
    \includegraphics[width=0.5\textwidth]{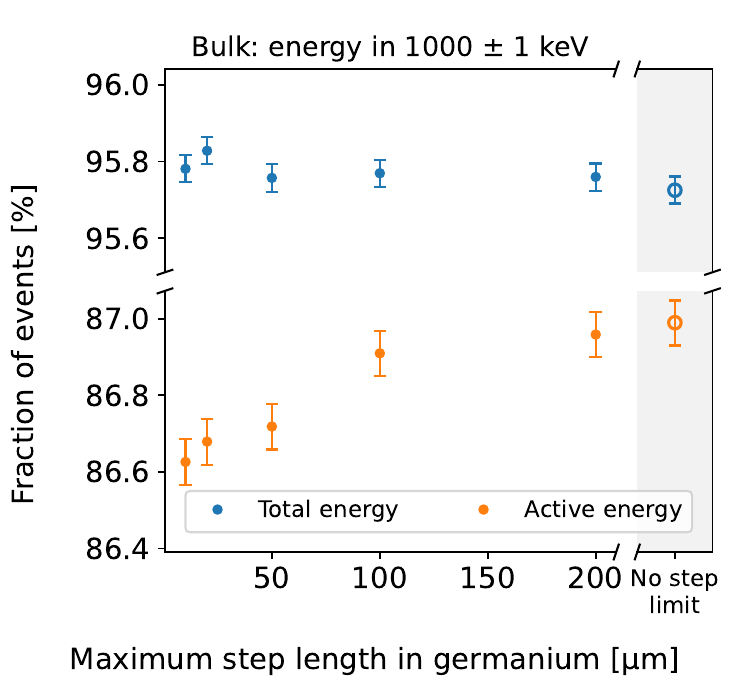}%
    \includegraphics[width=0.5\textwidth]{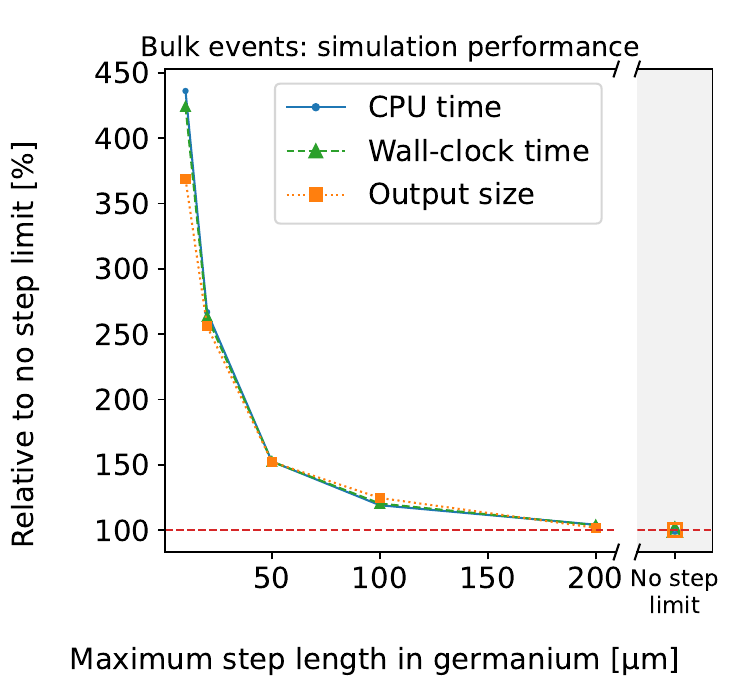}
    \caption{%
    Effect of step limits on simulated observables, for 1~MeV electrons generated throughout the full volume of a p-type germanium detector of typical geometry. The left panel shows the fraction of events whose collected energy, obtained by weighting each energy deposition by a charge collection efficiency that rises linearly with depth over a thickness typical of the detector surface region, falls within $1000\pm1$~keV, as a function of the maximum step length imposed in germanium, with the dashed line and band marking the no-step-limit result. The fraction decreases as the step length is shortened, and the attribution of energy depositions to the differently sensitive parts of the germanium volume gets more precise. The right panel shows the associated computing cost: CPU time, wall-clock time and output file size, each relative to the no-step-limit case, all grow steeply as the step length is shortened.}\label{fig:step-limits-validation}
\end{figure}

Production cuts and step limits can affect simulated observables, so the validation suite quantifies their effect, both to determine the defaults above and to guide users who choose their own. Because \geant always ends steps when reaching a geometrical boundary and also overrides the production cut for secondaries very close to boundaries, the energy deposit attributed to each volume is largely independent of the step length or production cut. Near the surface of a p-type \ac{HPGe} detector, however, the charge collection efficiency follows a continuous profile that is applied to the energy depositions during post-processing (\cref{sec:postproc,sec:validation}), and applying it correctly requires knowing where in this region each deposition occurred. \geant only reports the total energy deposited along a step, delimited by its start and end points, without resolving where along it the energy was released; \remage by default assigns the deposit to the midpoint between the two. The step is in any case not the true trajectory of an electron, for example, since multiple scattering is modeled only approximately~\cite{Ivanchenko:2010zz}. Because this region is not a boundary of the simulated geometry, \geant does not shorten steps there as it does when a track approaches one, and a shorter step length has to be imposed by the user. \Cref{fig:step-limits-validation} quantifies this for 1~MeV electrons generated throughout the full volume of a p-type germanium detector modeled with a cylindrical geometry, weighting each energy deposition by a charge collection efficiency that rises linearly with depth over a thickness typical of this region. Shortening the maximum step length from 200 to 10~$\mu$m lowers the fraction of events with the full collected energy by about 1\%, while the CPU time and the output size grow substantially. These numbers are only an example, and the trade-off between accuracy and performance needs to be judged for each simulation goal. The impact of the step length depends on the simulated physics and on the detector geometry, and in particular on the charge collection efficiency profile assumed near the detector surface.

\paragraph{Hadronic physics.}
Radiogenic background studies for low-background experiments usually require only electromagnetic physics. Hadronic physics becomes necessary for two further classes of background: the atmospheric muons that cross the detector, and neutrons, whether produced by those muons or released in the surrounding materials. Muons can induce a signal through their own ionization, through the particle showers they generate, or through the time-delayed decays of the many radioactive isotopes those showers produce; the last two require hadronic interactions to be modeled, as does the transport of neutrons and their capture on the surrounding nuclei.

In \remage, hadronic physics is activated through a macro command, with which the user also selects among several constructors modeled after the hadronic parts of \geant's reference physics lists (\hypentt{QGSP\_BIC\_HP}, \hypentt{QGSP\_BERT\_HP}, \hypentt{FTFP\_BERT\_HP} and \texttt{Shielding}).

All of them are initialized together with \geant's Neutron-HP model, which treats neutrons below 20~MeV with evaluated cross-section data, both for inelastic and elastic scattering (\hypentt{G4HadronElasticPhysicsHP}), and \geant's thermal neutron scattering can be enabled on top with an additional command (\hypentt{G4ThermalNeutrons}). \remage further simulates the capture of negatively charged particles coming to rest, muons among them (\hypentt{G4StoppingPhysics}), and the inelastic nuclear interactions of light and heavy ions (\hypentt{G4IonPhysics}).

When isotopes are produced by neutron capture, a good model of the corresponding $\gamma$ cascade is needed. \geant provides two built-in statistical models for generating these $\gamma$ rays: \hypentt{G4NeutronHPCapture}, based on the G4NDL library, and \hypentt{G4PhotonEvaporation}, based on level-scheme data from ENSDF. Neither is accurate enough to model the detector response with the precision a tagging strategy requires (see the discussion in~\cite{Weimer:2026}), so an alternative generator is required. \remage therefore allows the default de-excitation to be overridden with $\gamma$ cascades read from external input files, specified by isotope and by range of incoming neutron kinetic energy. The file format is based on the output of MAURINA~\cite{Grabmayr:2024faq,Grabmayr:2023tpj}, a Hauser-Feshbach generator that, unlike the two models above, also accounts for the initial spin distribution of the compound nucleus and for the level density and parity distribution during de-excitation. Keeping the cascades in separate files nevertheless makes the mechanism independent of the generator that produced them. Other generators, such as NuDEX~\cite{Mendoza:2023}, now part of \geant, and \hypentt{G4RiversideCASCADE}~\cite{Weimer:2026}, are candidates for native support in future \remage releases.

\begin{figure}
    \includegraphics[width=0.5\textwidth]{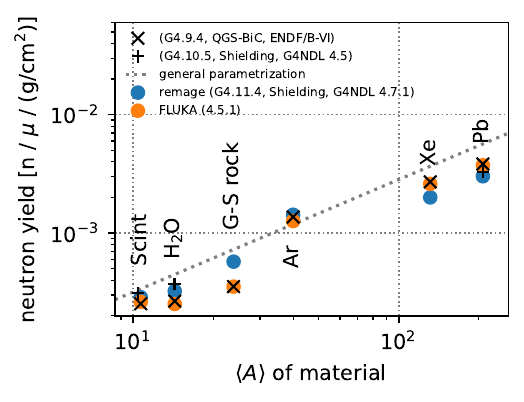}%
    \includegraphics[width=0.5\textwidth]{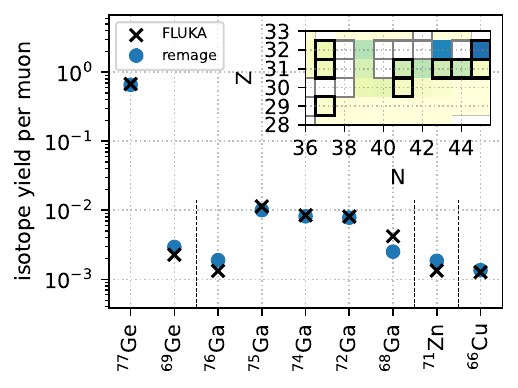}
    \caption{%
    Selected hadronic physics cross-checks from the \remage validation suite. The left plot shows the simulated muon-induced neutron yield for various materials used in low-background experiments with muons started at 280~GeV. It compares \remage with FLUKA, previous \geant-based studies~\cite{Manukovsky:2016fcn,Pec:2023yic} for precise comparison, and a general parameterization~\cite{Agafonova:2013bva} as a rough estimate. The right plot compares simulated muon-induced isotope production in germanium enriched to about 92\% in $^{76}$Ge using \remage and FLUKA, with muons starting at 100~GeV and identical geometry. The heat map in the inset shows the predicted FLUKA production rate and highlights the location of the selected isotopes in the nuclear chart. Differences in both plots compared to reference simulations can be due to systematic differences in models and potentially also to differences in the methodology used to estimate the observables. Therefore, only large deviations should be flagged.}\label{fig:muon-physics-cross-check}
\end{figure}

Because the background from the decays of muon-induced isotopes must be well understood, both the simulated isotope production rate and any accompanying signal that would allow the production to be tagged need to be validated. Validation tests on muon physics consist of two parts. The first concerns the mean energy loss per unit of slant depth of muons in different materials and at different kinetic energies. The second considers the mean particle yield in hadronic showers, specifically, of neutrons and spallation isotopes, for different materials, at different muon kinetic energies, and different physics lists. Where available, the simulation estimates are compared to experimental values, but for all combinations tested, parameterizations based on experimental and simulation data are employed. In addition, FLUKA simulations~\cite{Ahdida:2022gjl,Battistoni:2015epi,FLUKAWebsite} with the same setup were performed and are used as an additional reference. Due to the considerable uncertainties in both experimental and simulation estimates, deviations from these should be considered significant only when they are large. A selection of plots provided in the validation suite is shown in \cref{fig:muon-physics-cross-check}. The suite also covers the transport and capture of neutrons themselves, against evaluated nuclear data.

\paragraph{Optical physics.}
Many low-background experiments also read out the light that energy depositions produce through scintillation or Cherenkov emission. To simulate the response of these channels as well, the physics list can be extended with a macro command that enables the tracking of optical photons and their interactions. This part follows the upstream \geant physics closely, adding the two emission processes above (\texttt{G4Cerenkov} and \hypentt{G4Scintillation}) together with the optical boundary, bulk absorption and Rayleigh scattering processes (\hypentt{G4OpBoundaryProcess}, \hypentt{G4OpAbsorption} and \hypentt{G4OpRayleigh}), as well as the \ac{WLS} process in the modified form described below. The scintillation process is configured to distinguish the emission spectrum and yield of different particle types. Enabling these processes is not sufficient on its own, however. The optical properties of the individual detector materials must also be supplied by the user as part of the experimental geometry, as described in \cref{sec:geometry}.

Once a \ac{WLS} efficiency is assigned to a material, \geant's wavelength-shifting process (\texttt{G4OpWLS}) draws the number of shifted photons from a Poisson distribution. Even though this efficiency is below unity for the usual shifters, such as tetraphenyl butadiene, a single incident photon can still yield more than one shifted photon, and the total energy of the emitted photons is never compared against that of the incident one, so energy conservation is not guaranteed either. Neither behavior is appropriate for these materials.
By default, \remage therefore replaces this process with a custom one that emits at most one photon, performing a Bernoulli trial with the configured efficiency in place of the Poisson sampling. \geant's original process can be restored with a macro command. \remage additionally opts in to sampling the emission time from an exponential profile rather than from \geant's default delta function.

Optical physics is sensitive to physics or infrastructure changes, and the validation suite verifies the optical simulation microphysics. As an example, the validation suite checks whether the expected number of photons is generated in a scintillation or Cherenkov event, is absorbed after various lengths of absorbing materials, or is detected at a photo-sensor at the correct rate.

\section{Simulation execution and parallelization}%
\label{sec:running}
All user interaction with \remage happens through a single command-line entry point, the \texttt{remage} executable. A typical invocation only needs to specify a detector geometry, a macro file describing the simulation to run, and an output file; for example:

\begin{lstlisting}[frame=none, basicstyle=\ttfamily\footnotesize]
$ remage -g geometry.gdml -o output.lh5 [OPTIONS] macro.mac
\end{lstlisting}

This basic invocation runs the simulation in a single process. Simulations for low-background physics, however, often have to generate very large numbers of events, since background sources frequently have a low probability of producing a detectable signal. Techniques that bias the simulation toward these rare paths are often challenging to apply, so scaling the simulation up to large computing clusters is usually the most direct way to keep it tractable.

\remage therefore offers two mutually exclusive ways of parallelizing a simulation run, selected with dedicated command-line flags: a threading mode and a multi-processing mode. The threading mode delegates directly to \geant's own native multi-threaded run manager. A single \remage process spawns several worker threads that share the same memory space, so that objects common to the whole run, notably the geometry, are built only once regardless of how many threads are used. The multi-processing mode is instead implemented by \remage itself, independently of \geant. The executable launches several independent processes, each running its own complete, separate \remage instance with its own private memory.

In both modes, each worker thread or process writes its own output file, identified by a thread- or process-specific suffix; these partial files can optionally be merged back into a single file once the run completes.

\begin{figure}
    \centering
    \includegraphics[width=\textwidth]{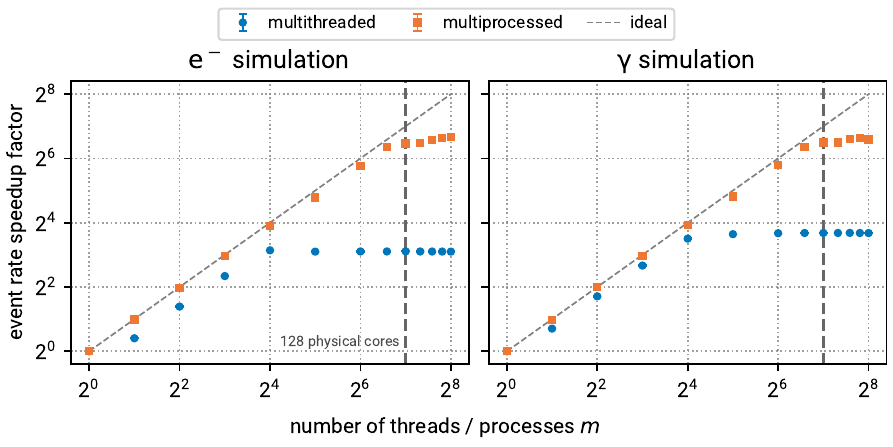}
    \caption{%
      The speedup factor for simulated event rates as a function of the number of parallel processes or threads. The setup used a geometry consisting of a germanium box in liquid argon, both registered as sensitive volumes. The left panel used primary electrons, and the right panel used gammas, both started with the general particle source (GPS) at 1~MeV without confinement. Each process or thread used the same number of primaries (testing weak scaling). Simulations ran on nodes with two AMD EPYC 7763 CPUs (128 physical cores) and saved output to flash storage with 204.8~GB/s bandwidth. In both cases, performance scaled linearly at first, then slowed or saturated. Multi-processing slowed down at around the number of physical cores due to entering the hyper-threaded regime. Multi-threading saturated earlier, and the saturation point depends on the setup and primary particle.
    \label{fig:performance-scaling}}
\end{figure}

The two modes differ most importantly in the following respects. Because every process in multi-processing mode carries its own full, independent memory footprint, this mode is markedly more resource-hungry than the threading mode, where sharing the geometry and physics tables keeps the additional memory cost of each worker thread low. The two modes also scale differently with the number of parallel workers. Both scale linearly at first, though the threading mode usually with a lower slope, and both eventually saturate. In multi-processing mode this saturation sets in around the number of available physical CPU cores; in threading mode it is reached earlier. This behavior is shown quantitatively in \cref{fig:performance-scaling}, for two different sets of primary event setups in the same geometry. The source of this congestion in multi-threaded mode is not fully understood, and it also is dependent on the choice of simulation setup.

Because the threading mode relies directly on \geant's own multi-threading implementation, this behavior reflects the \geant version used at the time of writing and may change as that implementation evolves. As a rule of thumb, the threading mode is the better choice when memory is the limiting resource, while the multi-processing mode gives better overall throughput otherwise. Both modes are also fully compatible with externally supplied primary events (\cref{sec:vertex,sec:generators}).

\section{Output and data reduction}%
\label{sec:output}

Users of simulation tools like \remage need the outcome of the simulated particle interactions written to disk in a form suitable for further analysis: information about recorded interactions of interest in a detector must be captured in a structured form. \remage stores this information in table-like structures, arranged hierarchically within the output file much like a directory tree. What is written, and how it is organized, is governed by a set of configurable output schemes (\cref{sec:outputschemes}).

\remage's primary output format is \ac{HDF5}~\cite{HDFGroup:2025}, following the LEGEND \acs{HDF5} specification~\cite{legend-data-format-specs_2020}; only this format exposes \remage's most advanced output features, described below. \remage also supports all the other output formats provided by \geant's own \texttt{G4AnalysisManager} (plain \ac{HDF5}, ROOT, CSV, and others), and the same set of formats is used for reading in primary particle information, as already discussed in \cref{sec:vertex,sec:generators}. \remage's \acs{HDF5} output can be read straightforwardly with either the Python or Julia packages from the LEGEND software ecosystem~\cite{legend-lh5io_2026,LegendHDF5IOjl_2018}. However, \ac{HDF5} is standardized and commonly used in many other scientific applications, so the files can additionally be used with any of the many other libraries capable of opening \ac{HDF5} files. These output files are also compressed with the Zstandard algorithm~\cite{RFC8878} to reduce disk space usage.

What is written and in which format is only part of the picture. Depending on the physics processes and settings enabled, a simulation can record more detail than is actually needed, and can spend time tracking particles that never affect the result. \remage offers measures against both, targeting the detailed microscopic simulation inside active detectors (\cref{sec:clustering}) and the simulation of particles that do not contribute to the final observables (\cref{sec:filtering,sec:conditional}).

\subsection{Output schemes}%
\label{sec:outputschemes}

The content and structure of \remage's output are determined by what we call ``output schemes'': interchangeable components that implement event selection, the persistency of simulated quantities to disk, and, for some of them, control over which particles are tracked (\geant's ``stacking action''). Schemes fall into two categories: those tied to a detector type and those global to the run. The former relate directly to the ``detector'' concept already introduced in \cref{sec:detectors}. Every detector has a type, one of \texttt{Germanium}, \texttt{Scintillator}, \texttt{Optical} or \texttt{Calorimeter}, and a unique identifier chosen by the user, either through a macro command or in the \ac{GDML} file itself, as described in \cref{sec:detectors}. Registering any sensitive volume of a given type automatically activates the corresponding output scheme, which records only interactions occurring inside the registered volumes. The macro command selects volumes by regular expression and assigns identifiers that are either incremented automatically across the matched volumes or shared by all of them, so that a whole array of detectors can be registered in a single line. Volumes sharing an identifier are written to the same output table, and are thus essentially handled as one large detector volume.

The observables stored per hit or event depend on the scheme, but typically include the interaction time, particle type, position and deposited energy. The number of quantities written out in a simulation not only affects the storage space required, but also strongly affects the overall runtime of the simulation. Encoding and storing output data to disk can take a substantial amount of simulation time; in some cases, the simulation performance can even be limited by the I/O throughput of the used storage device. While in the past (such as in the output classes implemented in \mage) often most tracking information was written out for later post-processing, \remage takes a different approach: it minimizes this amount by selecting only the minimum necessary set of quantities stored in the tabular output data. All output schemes tied to detector types try to restrict their output in this way, but often give users additional flexibility to selectively turn on other data columns if required for their analysis.
This reduction complements the other strategies to minimize the amount of output data in \remage that are described in subsequent sections.

\remage currently offers the following detector-specific output schemes:
\begin{itemize}
  \item \texttt{Germanium}: records step-wise, rather than summed, hit information, including the distance of each step from the detector surface; recommended for \ac{HPGe} and other solid-state detectors, useful for offline pulse-shape and active-volume modeling.
  \item \texttt{Scintillator}: follows the same step-wise philosophy as \texttt{Germanium}; recommended for scintillating or other bulk media, such as liquid argon, useful for offline optical-response modeling.
  \item \texttt{Optical}: records the timestamp and wavelength of optical photons detected at the detector surface, terminating their tracks there; recommended for optical detectors like \acp{SiPM} and \acp{PMT}.
  \item \texttt{Calorimeter}: a minimal scheme that stores, for each \geant event, only the total deposited energy and the time of the first energy deposit, without any per-step detail; recommended for simple veto or absorber detectors.
\end{itemize}

The second category comprises three further schemes: \texttt{Vertex}, which records the primary-vertex information of every simulated event and is activated automatically whenever any other output scheme is active; \texttt{Track}, an optional scheme that stores the initial state of every simulated track; and \hypentt{GeometryCheck}, already described in \cref{sec:geometry}, opt-in like \texttt{Track} but targeting geometry validation rather than detector output.

The step-level detail recorded by the \texttt{Germanium} and \texttt{Scintillator} schemes lets downstream post-processing software, such as \pkgformat{reboost}~\cite{reboost_2025} from the \remage ecosystem, apply detector response models offline, without repeating the computationally expensive tracking simulation, a two-stage approach commonly adopted by experiments in the field~\cite{Agostini:2010rd,Carrettoni:2011rn,GERDA:2022hxs,LZ:2020zog}. Such detail is required because the response of these detectors depends on where the energy was deposited and not only on how much: an \ac{HPGe} detector is sensitive to the spatial topology of the depositions through the resulting pulse shape, while in a scintillating volume the probability that the emitted light reaches an optical detector varies with the position of the deposition. Where this level of detail is not needed, a volume can instead be registered as a \texttt{Calorimeter}.

In an actual experiment, an event is defined as the set of signals falling within a time window set by the detector's time resolution, and \remage aims to reproduce this definition for simulated events. The \geant event index alone does not achieve this. A \geant event describing a full radioactive decay chain (\cref{sec:physicslist}), for instance, can include energy depositions from a delayed daughter decay occurring long after those from the parent, well beyond the time window of any real detector, yet still attributed to the same \geant event.

\remage therefore introduces the concept of a ``hit'': the set of steps recorded within a single detector that fall within a given time-coincidence window. A new hit begins whenever the underlying \geant event index changes, or whenever, within the same event, the recorded time jumps by more than the window size since the previous step. The window defaults to 10~$\mu\mathrm{s}$, a value driven by the typical time resolution with which \ac{HPGe} detectors separate two signals into distinct hits, and can be adjusted by the user where a different value is needed. As a result, one row of a \remage output table does not necessarily correspond one-to-one with one simulated \geant event; in the decay-chain example above, \remage splits the single \geant event into several separate hits.

For the step-wise schemes, \texttt{Germanium} and \texttt{Scintillator}, the number of steps in a hit varies, so the columns holding step-level quantities are stored as variable-length arrays rather than as fixed-width columns. This is the natural representation for data of this kind, and it simplifies the analysis, since all the steps belonging to a hit are already grouped together. Such variable-length columns are part of the LEGEND \ac{HDF5} specification~\cite{legend-data-format-specs_2020} and are not available in the other supported formats.

\begin{figure}
    \centering
    \includegraphics[width=\textwidth]{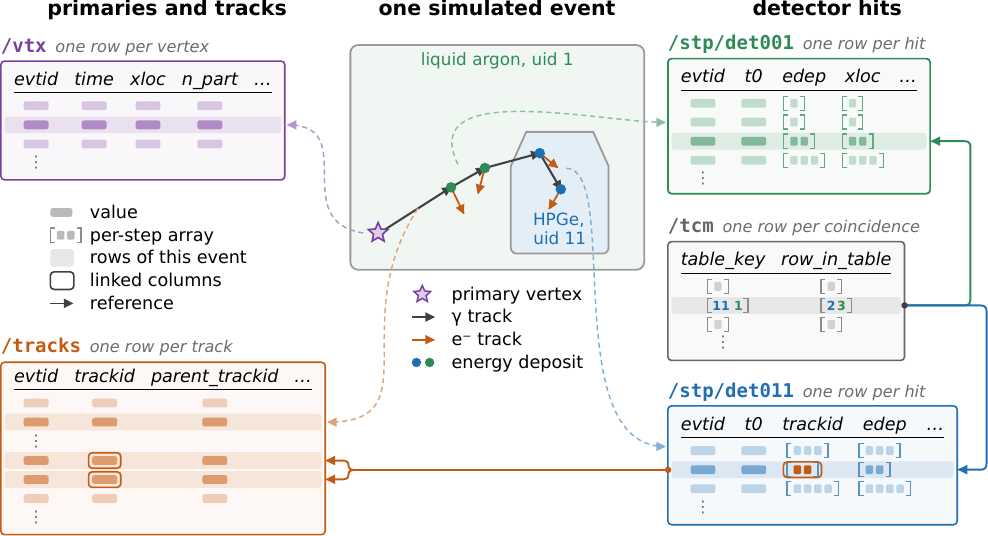}
    \caption{%
      Simplified sketch of \remage's output table structure in an \acs{HDF5} file. In this simple example, a $\gamma$-ray is started in a liquid argon volume and deposits energy both in this scintillator volume and in a germanium detector.
      The \texttt{/vtx} and \texttt{/particle} tables contain information on the primary $\gamma$-ray, e.g., its position, energy and momentum. Similar information is (optionally) also stored for all secondary tracks in the \texttt{/tracks} table.
      Actual detector energy depositions are stored in tables below \texttt{/stp/} labeled by the unique detector ID number (uid). These are also reshaped in such a way that depositions close together in time are not stored in separate rows but in variable-length fields as a ``hit''. Even in this simple two-detector setup, the information from this one \geant event is already scattered across two detector tables. To ease iterating over these for subsequent analysis, the time-coincidence map in \texttt{/tcm} stores references to the rows in the individual detector tables that fall into the same time window.
    \label{fig:output-model}}
\end{figure}

Two further features concern how the hits are distributed across tables. By default, each detector is written to its own output table, but the user can instead request a single, merged table per detector type, with the original detector volume labeled by the numeric detector ID. Since the default layout scatters the hits of a single event over several tables, \remage also computes and stores in the output file a lookup table, the \ac{TCM}, that lets downstream analysis reconstruct which rows in the different detector tables belong to the same time-coincident event, without having to scan every table.
\Cref{fig:output-model} showcases the interplay of all described tables using the example of a germanium detector in argon.

\subsection{Step clustering}%
\label{sec:clustering}

The step-wise detail recorded by the \texttt{Germanium} and \texttt{Scintillator} schemes can substantially increase the output volume and the associated performance degradation due to high system I/O load. Often, however, \geant takes steps much shorter than what is meaningful in an \ac{HPGe} or a scintillation detector, well below the experimental resolution. In germanium, for example, the charge clouds produced by interactions are typically 1--2~mm across, so the simulation is not sensitive to tracking at the micrometer level. \remage therefore clusters steps that lie close together, reducing the output file size while retaining the information needed to compute the observables of interest.

\begin{figure}
    \centering
    \includegraphics[width=\textwidth]{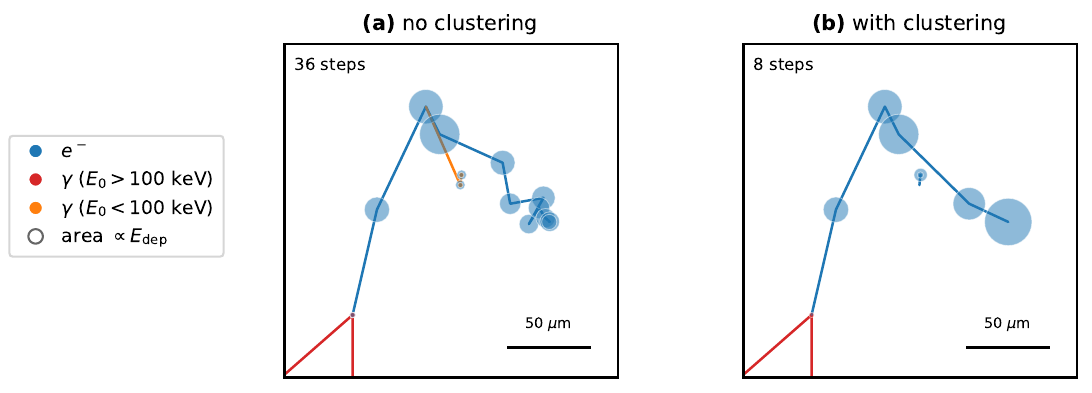}
    \caption{%
      Microscopic view of simulated steps in a germanium volume, after Compton scattering of a $\gamma$-ray (bottom left). Without clustering (left panel), many small steps of the secondary electrons, as well as secondary, low-energy bremsstrahlung $\gamma$s are visible. The post-step points of all particles' steps are indicated with points whose area is proportional to the energy deposition in the step leading to this point. With clustering (\remage's default settings, right panel), these low-energy energy depositions are merged into fewer steps, each with a higher energy. The clustering preserves the location of the energy depositions. The number of secondary tracks or their particle type are not necessarily preserved. For example, the bremsstrahlung track is merged into one of the electron tracks.
    \label{fig:clustering}}
\end{figure}

Clustering is restricted to steps within the same \texttt{G4Track}, so that each cluster remains interpretable as an effective step in the detector, combining the time, position and energy of its constituent steps, simply with a larger step length. The output format is unchanged: clustering reduces the number of rows written, not their meaning. One exception is made for very low-energy tracks: some \geant processes, such as atomic de-excitation, produce many secondary tracks with very low energy and range that are not expected to affect the observables of interest. Each such low-energy track is merged into the neighboring track with the highest energy among those whose first pre-step point lies within the same distance threshold of its own first pre-step point. After this pre-processing step, \remage walks through the steps of each track, joining a step to the current cluster whenever its distance from the cluster's first step and its time difference from that step both stay below user-defined thresholds; this clustering time threshold is unrelated to the time-coincidence window used to define hits (\cref{sec:outputschemes}). Germanium detectors require finer treatment near their surface, where the charge collection efficiency varies steeply with depth, and spatial detail that is irrelevant in the bulk directly affects the modeled response there. A separate clustering distance threshold therefore applies to the surface region, and clusters are broken whenever a track crosses the boundary between the two regions. \Cref{fig:clustering} illustrates the effect of this clustering.

Clustering is enabled by default and fully customizable through macro commands. The default thresholds are chosen to stay well below the resolution any real detector achieves, so that no physically useful information is discarded, while still removing a large fraction of \geant steps. For \texttt{Germanium}, steps are clustered when they lie within 50~$\mu\mathrm{m}$ and 10~$\mu\mathrm{s}$ of each other, and the merging of low-energy tracks described above applies below 10~keV. Within the 2~mm region closest to the detector surface, where an \ac{HPGe} detector's response is most position-sensitive, no clustering is applied. These defaults represent a deliberately minimal first pass at data reduction, leaving the choice of a more aggressive reduction, where needed, to the user.

\begin{figure}
    \centering
    \includegraphics[width=0.5\textwidth]{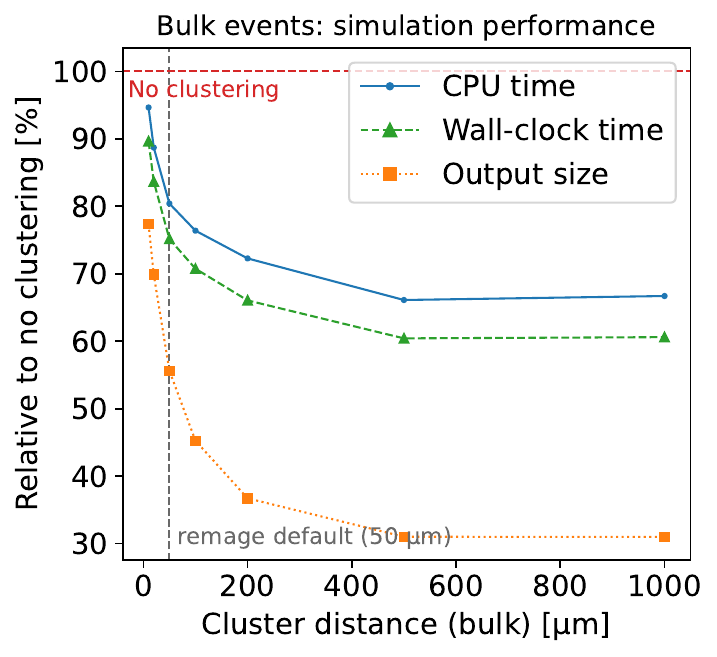}
    \caption{%
     Performance impact of step clustering, for a setup with 1~MeV electrons generated uniformly in the bulk of a germanium detector. Compared to the case without step clustering, both the CPU time spent and the output size are lowered significantly.
    \label{fig:clustering-performance}}
\end{figure}

By reducing the number of step-wise rows written to the output tables, clustering lowers both the output file size and the associated I/O cost. \Cref{fig:clustering-performance} shows the performance gains achieved by \remage when step clustering is enabled. It is a simple simulation of 1~MeV electrons in the bulk of a germanium detector that is registered as a sensitive volume. Even small clustering thresholds (such as \remage's default) yield a significant reduction of the simulation's CPU time and output size.
Not shown is the further performance gain in the post-processing steps. With less data to process, both the hit-like reshaping (see \cref{sec:outputschemes}) and the users' analysis code require fewer computing resources.
This performance gain increases for larger clustering radii, so the radius needs to be chosen by looking at its impact on the simulated physics (not shown here).

\subsection{Event filtering}%
\label{sec:filtering}

Event filtering decides, for each event, whether it is worth writing to disk; it acts purely at the output stage and leaves the simulation itself unaffected. It is useful whenever a large fraction of the simulated events could be irrelevant to the subsequent analysis. \remage offers two criteria for this decision. The first is the energy deposited, available for \texttt{Germanium} and \texttt{Scintillator} detectors: the user sets upper and lower thresholds and selects which detectors participate, and the event is written only if the summed energy deposition across them falls within the bounds. This can be used to reproduce in the simulation the trigger energy threshold of the real experiment. The second is the production of a given isotope: a dedicated filtering scheme keeps only those events in which a specific isotope has been produced. This is useful, for instance, in simulations of rare neutron activation, where only a small fraction of the simulated events produce the isotope of interest.

\subsection{Conditional tracking}%
\label{sec:conditional}

Simulating optical photons in \geant is computationally expensive, and a typical scintillating medium produces a large number of them in a single event. Reading out such a medium as a veto is a common experimental technique, and simulating it faithfully means tracking the scintillation light all the way to the optical detectors that collect it. In such a setup the decision on whether an event is worth keeping at all rests on a different detector. In LEGEND, for example, one of the triggering schemes is provided by the \ac{HPGe} detectors, while the surrounding liquid argon is read out as a veto. If no energy is deposited in the germanium, the event is discarded regardless of what happens in the argon, so the effort spent tracking its optical photons is wasted. Conditional tracking exists to avoid exactly this. The energy criterion that decides the fate of the deferred tracks is available only for \texttt{Germanium} detectors.

Deferred tracks are not tracked immediately but stored for later. Optical photons are the natural candidates, as deferring them leaves the physics untouched; they cannot alter the energy deposition on which the later decision for tracking them is based.
Electrons can be deferred as well, and optionally positrons under the same conditions; for them the user gives an energy threshold and a minimum distance from the \texttt{Germanium} detector surfaces, chosen so that the discard decision described above is not itself affected by the deferral. The distance matters because an electron need not reach a detector for its byproducts to do so, bremsstrahlung photons being one example, and these can deposit energy well away from where the electron itself stops. Depending on whether the conditions for further tracking are satisfied, the stored tracks are either tracked or dropped along with the rest of the event. The energy entering that decision is summed over all \texttt{Germanium} detectors, or over a user-chosen subset of them. The deferral acts at the moment a track is created; a companion mechanism, suspension, acts during tracking instead, halting an electron whose energy drops below a configured threshold and deferring it in the same way.

Conditional tracking is not free of consequences for the physics. Deferring electrons is more delicate than deferring optical photons alone, since electrons can carry a significant fraction of the energy in an electromagnetic cascade. Part of an energy deposit may be carried by the deferred tracks themselves and therefore go uncounted when the discard decision is taken, so that the condition fails and the waiting tracks are dropped, and deferring high-energy shower components can likewise suppress isotope production. It should therefore be validated against the physics observables of interest before being relied upon.

Beyond \geant's own staging mechanism, \remage adds a custom deferral mechanism, because optical simulations can leave a vast number of waiting tracks that occupy considerable memory; it spills waiting tracks from memory to a scratch file, with negligible impact on processing speed, as validated with the \remage validation suite.

As an example, two deferral configurations were tested using the geometry from the speedup tests in \cref{fig:performance-scaling}, with 2.6~MeV gammas sampled 50~cm from the \ac{HPGe} and a 100~keV minimum energy threshold in the \ac{HPGe}. Deferring only the optical photons gave a 5$\times$ speedup, while deferring both optical photons and electrons more than 10~cm from the \ac{HPGe} yielded a 170$\times$ speedup. The greater speedup is due to avoiding the overhead of placing optical photons on the waiting stack by deferring their parent electrons as well.

\subsection{Particle filtering}%
\label{sec:particlefilter}

The particle filter follows a similar but more drastic approach: rather than deciding after the fact whether an event is worth writing to disk, it removes selected particles from the simulation, immediately after they are created and before they are tracked, with no regard for energy conservation. This differs from \geant's production cuts (\cref{sec:physicslist}), which stop a particle only once its remaining range falls below a threshold and deposit its energy locally to conserve the event's energy budget.

This can reduce simulation effort by discarding particles with a short range produced far from any detector and therefore unable to contribute to the observables of interest, such as $\alpha$ particles from radioactive decay in a $\gamma$-ray experiment, or serve to verify that a given particle species does not contribute to the output, by removing it and comparing the result. The user specifies which particle species the filter targets, and can further restrict its action to chosen production processes, chosen production volumes, or both, in each case configuring the filter to keep only the matching particles or to remove only them. Because it breaks energy conservation, results obtained with the particle filter call for the same caution as any deliberately incomplete physical picture.

\subsection{Processing the simulation output}%
\label{sec:postproc}
The \remage output can be analyzed as it is, but an optional post-processing step allows the response of the detectors to be applied to it afterwards. \pkgformat{reboost}~\cite{reboost_2025} implements this step as a collection of processors that act on the \remage output, so that the same simulation can be reprocessed under different response models without repeating the tracking. Among them are the activeness model that weights each energy deposit by its distance from the detector surface, and a range of models for observables related to the shape of the detector signal, from simple heuristics based on the spatial spread of the energy deposits or on their drift time, to an analytic model of the current pulse and to libraries of simulated pulse shapes. Optical maps can also be applied, replacing the tracking of individual scintillation photons with a precomputed, position-dependent detection probability for each optical channel; \pkgformat{reboost} provides the routines to produce such maps from a dedicated optical simulation as well. Others reproduce the response of the readout, from the smearing of the energy with the detector resolution to the trigger and pile-up behavior of the data acquisition system. Another typical workflow for \ac{HPGe} detectors is worth a mention: the simulated energy depositions are passed to a dedicated signal simulation tool, such as SolidStateDetectors.jl~\cite{Abt:2021mzq}, and the language-agnostic \ac{HDF5} output makes this step seamless.

\section{End-to-end validation with a germanium detector and a radioactive source}%
\label{sec:validation}

The validation tests described so far are component-level: each isolates a single algorithm, physics process or generator and checks it against a theoretical expectation, a reference database or an independent calculation. To also test the full simulation chain against an actual measurement, we reuse one of the existing characterization runs~\cite{hades-th228-spectra} that the GERDA detectors underwent before deployment~\cite{GERDA:2019vry}, performed at the \ac{HPGe} detector characterization facility in the HADES underground laboratory~\cite{Andreotti:2013xoa}. In this 17-hour run, a BEGe detector was exposed to a $^{228}$Th point source of about 6.3~kBq, placed on the detector symmetry axis 41~mm above the detector face. Quality cuts against pile-up and unphysical events retain about 84\% of the recorded events.

\begin{figure}
    \centering
    \includegraphics[width=\textwidth]{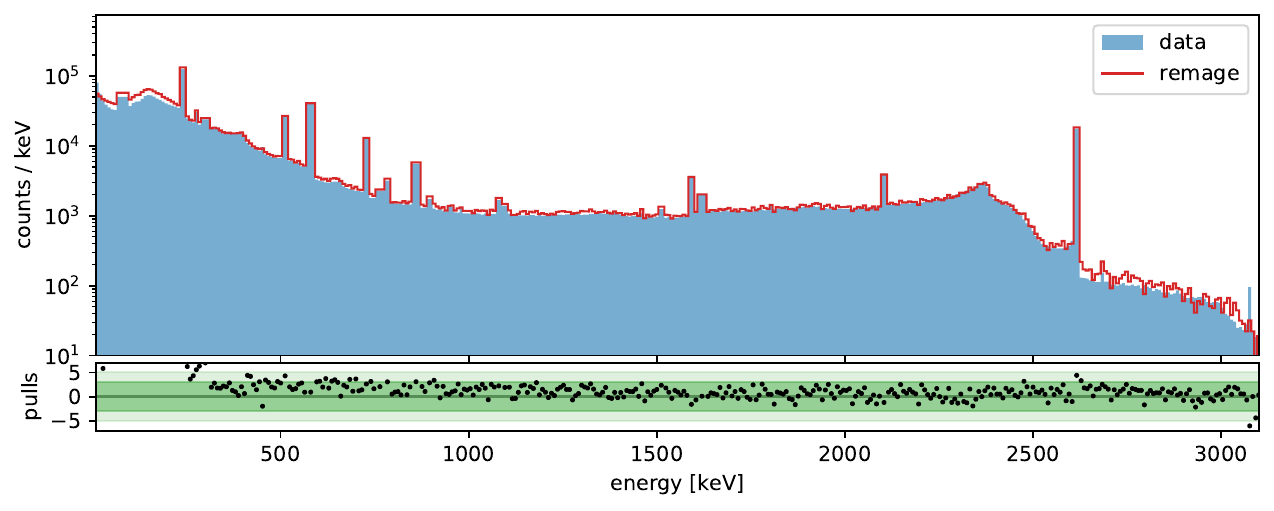}
    \caption{Comparison of a germanium energy spectrum simulated with \remage (red) to data (blue) taken by the GERDA collaboration at the \acs*{HPGe} detector characterization setup in HADES~\cite{Andreotti:2013xoa}, following exposure to a $^{228}$Th calibration source. The lower panel shows the residuals between data and simulation in units of standard deviations.}\label{fig:validation-spectrum}
\end{figure}

The corresponding simulation reproduces this setup, with the geometry built with \pkgformat{legend-pygeom-hades}~\cite{legend-pygeom-hades_v021}, and the $^{228}$Th decay chain generated with the \geant general particle source confined to the source volume, using \remage's default physics list and production cuts, without any tuning specific to this test. The resulting energy depositions are post-processed with \pkgformat{reboost}, applying a simple activeness model near the detector's n\textsuperscript{+} contact and a Gaussian energy resolution smearing to obtain a realistic-looking spectrum; the residual environmental radioactivity of the setup, measured in a separate run without the source, is rescaled to the live time of the characterization run and added to the simulated spectrum before the comparison. The simulated spectrum itself is scaled by the source activity, the live and dead time of the run, and the efficiency of the quality cuts. \Cref{fig:validation-spectrum} shows the resulting comparison, using a variable binning that is coarser around the $\gamma$-lines so that the result is not sensitive to their detailed peak shape.

Over most of the spectrum, above roughly 300~keV, the residuals between data and simulation stay within about 3 standard deviations, with a handful of points at lower energy showing larger deviations. Several sources of uncertainty are expected to contribute at this level, none of which concerns the particle transport itself: the simplified description of the detector geometry and of the surrounding materials, the knowledge of the source geometry and position, the uncertainty on the source activity and the approximate model of the detector active volume, whose effect is most pronounced at low energy. Given that the simulation uses \remage's default physics settings, with nothing tuned to reproduce this particular measurement, the level of reproduction achieved is good.

\section{Software development, documentation and distribution}\label{sec:development}
\remage is developed openly on GitHub~\cite{remage-repo} and released, like the surrounding ecosystem of Python packages, under the GPL-3.0 license. Development follows modern software engineering practices, centered on \ac{CI/CD}. Every proposed change is automatically built and tested, and the resulting artifacts, from container images to documentation, are published automatically rather than assembled by hand at release time. Code formatting and static analysis, for both C++ and Python, are enforced automatically on every change.

Continuous deployment extends to the user-facing documentation~\cite{remage-docs}, which is rebuilt and published on every change to the repository. The macro command reference introduced in \cref{sec:design} is generated automatically from the source code that defines the commands, so it cannot fall out of date.

Making \remage easy to install is treated as seriously as making it easy to use (\cref{sec:design}). Container images are published on Docker Hub~\cite{dockerhub}. \remage is also packaged for conda-forge~\cite{conda-forge}, for Linux and macOS for all common CPU architectures, and is therefore installable with conda~\cite{conda} or Pixi~\cite{pixi}. The surrounding ecosystem of Python packages, including \pkgformat{reboost}, \pkgformat{revertex}, \pkgformat{legend-lh5io} and the \pkgformat{legend-pygeom} family of geometry packages, is available separately on PyPI and conda-forge. Together, these channels cover the full simulation stack, from the compiled \remage application to the surrounding Python tooling. Every release of \remage and of the ecosystem packages is additionally archived on Zenodo, which assigns it a citable \ac{DOI}.

For each \remage release we recommend a specific \geant version, which is also the one our distributed binaries and container images are built against. We provide compatibility with new \geant releases as quickly as we can and test against them immediately, but the version we recommend is the one that has passed the validation suite. Since the validation simulations are far more demanding than the ordinary tests, this suite runs on a separate machine that we maintain ourselves, with considerably more computing power than the machines the CI service provides.

The \texttt{remage} executable that users invoke is itself a thin Python wrapper around the compiled C++ application. It pre-processes the input, orchestrates the underlying C++ processes, which is what makes the multi-process parallelization mode of \cref{sec:running} possible, and post-processes the output into the primary \ac{HDF5} format. The two sides communicate over Unix pipes, which also carry requests from C++ for Python to handle, such as validating the \ac{GDML} input. Keeping this outer layer in Python lets us extend what \remage can do without writing new C++ code. The few users whose needs go beyond the macro command interface can instead work from the other end: \remage is also installed as a library, so they can link against it and plug in their own C++ components, reusing everything else.

\section{Applications}%
\label{sec:usage}

LEGEND is moving its entire simulation effort to \remage, away from the previous \mage toolkit. LEGEND-200~\cite{LEGEND:2025jwu}, which is already taking data, is simulated with \remage from a geometry built in Python with the \pkgformat{legend-pygeom} packages~\cite{legend-pygeom-l200_2024}. This is a complete experimental geometry, and one in which the argon scintillation light also has to be produced and propagated, so it exercises parts of \remage that a simpler setup does not. Nor is it a single, static description: it is designed to reproduce the multiple real configurations in which the experiment has run. LEGEND-1000~\cite{LEGEND:2021bnm}, the planned successor, uses the same software stack in the opposite mode~\cite{legend-pygeom-l1000_2026}: since the experiment is not yet built, its geometry is instead driven by design parameters that are varied to compare options, supporting design studies rather than reproducing an installed configuration. Two components of this ecosystem are not specific to LEGEND and can be used in any experiment of this kind. \pkgformat{legend-pygeom-hpges}~\cite{legend-pygeom-hpges_2024} simplifies the construction of \ac{HPGe} detectors in the geometry, building the types in common use from a standardized set of geometrical measurements. \pkgformat{legend-pygeom-optics}~\cite{legend-pygeom-optics_2024} is a catalog of the optical properties of the materials these experiments use, together with the methods to attach them to a geometry or to override them, which removes most of the bookkeeping that these many tabulated quantities otherwise require; its documentation is generated automatically and plots every property together with the reference it comes from, so the catalog can also be browsed online~\cite{legend-pygeom-optics-docs}.

A number of smaller test stands and $\gamma$-ray spectroscopy setups, with considerably simpler geometries, also rely on \remage. These include the LEGEND \ac{HPGe} detector characterization facility at HADES~\cite{Andreotti:2013xoa}, whose geometry package, \pkgformat{legend-pygeom-hades}~\cite{legend-pygeom-hades_2026}, is also the basis for the end-to-end validation of \cref{sec:validation}, and SCARF~\cite{Schwarz:2025nnq}, a liquid argon cryostat at TUM hosting \ac{HPGe} detectors for LEGEND-related R\&D, supported by the \pkgformat{pygeom-scarf} package~\cite{pygeom-scarf_2026}. \remage has also been used for the GeMPI radiopurity screening facility at LNGS and for BUGS~\cite{Scovell:2017srl}, the screening facility at the Boulby underground laboratory.

\section{Summary and outlook}%
\label{sec:conclusions}

This paper presented \remage, a framework designed to address the recurring challenges of running Monte Carlo simulations for low-background physics experiments. Its architecture separates the experimental geometry, built and checked with dedicated Python tools, from the core \geant application. \remage provides a physics list and event generators tuned to typical radiogenic and cosmogenic backgrounds, output schemes optimized for the most common detector types and analysis needs, and a validation suite that also quantifies the impact of key simulation settings on the observables of interest. It scales efficiently with the available computing resources, which keeps demanding simulations tractable, and further cuts computing cost directly through mechanisms such as step clustering, filtering and conditional tracking, which avoid spending effort on events or tracks that cannot affect the result.

\remage has become the simulation framework of the LEGEND experiment, and will therefore be developed and maintained for many years to come. Sustaining it is made easier by modern software development practices, such as continuous integration and deployment, automatically generated documentation, and fully open development, which also let its validation suite track new \geant releases while keeping the physics output stable. The feature set described here reflects \remage's state at the time of writing: development continues, the validation suite grows with it as new physics processes and detector types are covered, and the online documentation remains the authoritative, up-to-date reference.

For these reasons, we argue that \remage is a valuable resource, especially for small and medium-sized low-background experiments, since it already provides a high-quality, validated implementation of the features most of them need, without requiring each collaboration to develop and maintain its own. Through LEGEND and the several smaller, connected experimental efforts that already rely on it, \remage is also a resource for the wider low-background physics community.

\acknowledgments

\remage is developed within, and primarily for, the LEGEND collaboration, and much of what is described in this paper reflects that context. We thank the collaboration for its continued support of the software. We would like to thank the GERDA collaboration for the \ac{HPGe} characterization data used in the validation, and the \pkgformat{pygama} authors for the processing of those data~\cite{hades-th228-spectra}. We also thank the authors of \pkgformat{legend-pygeom-hades}~\cite{legend-pygeom-hades_2026} and \pkgformat{reboost}~\cite{reboost_2025} for the geometry implementation of that setup and for the post-processing tools used there.

This research is funded in part by the Deutsche Forschungsgemeinschaft (DFG, German
Research Foundation) -- Excellence Cluster ORIGINS EXC 2094-390783311; SFB1258-283604770.
We acknowledge the support of the German Federal Ministry of Research, Technology and Space (BMFTR) through
grant numbers 05A2023 and 05A2026.
This work has been supported by the Science and Technology Facilities Council (STFC), part of U.K.
Research and Innovation (grant number ST/W00058X/1).
This research is supported by the U.S. Department of Energy, Office of Science, Office of
Nuclear Physics under DE-FG02-97ER41041 and DE-FG02-97ER41033.
This research used resources
provided by the National Energy Research Scientific Computing Center (NERSC), a U.S. Department of
Energy Office of Science User Facility at LBNL.

Generative \ac{AI} tools were used at a later stage of the development of \remage, once the software had already reached an advanced and largely mature state; the great majority of the codebase was written by the human authors. Such tools were also used to assist in editing this manuscript.

\section*{Data and code availability}

\remage and all the ecosystem packages discussed in this work are openly developed on GitHub, in the \texttt{legend-exp} organization (\url{https://github.com/legend-exp}). Everything is released under the GPL-3.0 license. Every release of each package is archived on Zenodo and can be cited through its \ac{DOI}~\cite{remage,legend-pygeom-tools_2026}; this paper describes \remage v1.2.0~\cite{remage_v12} and \pkgformat{legend-pygeom-tools} v0.7.1~\cite{legend-pygeom-tools_2026_v071}. Ready-to-use conda-forge packages and Docker Hub container images are described in \cref{sec:development}, and the user documentation is available online~\cite{remage-docs}.

\bibliographystyle{JHEP-rmg}
\bibliography{bib/biblio.bib,bib/other-software.bib,bib/own-software.bib}

\appendix
\section{A complete simulation example}%
\label{app:example}

The two listings below make up a minimal but complete \remage setup. \Cref{lst:pygeom-example} builds the experimental geometry in Python and writes it to a \ac{GDML} file, following the workflow of \cref{sec:geometry}. \Cref{lst:remage-macro-example} is the macro file that runs the simulation on that geometry, registering the detectors, confining the event vertices and selecting the event kinematics.

\begin{lstlisting}[style=python, caption={Registering a detector with \pkgformat{legend-pygeom-tools}, attaching visualization and region metadata, and inspecting the registered detectors before writing the final \ac{GDML} file.}, label={lst:pygeom-example}]
import pint
from pygeomtools import Region, RemageDetectorInfo, write_pygeom
from pygeomtools.detectors import walk_detectors

u = pint.get_application_registry()

# reg, det_pv and det_lv are, respectively, a pyg4ometry.geant4.Registry
# and the physical and logical volume of the detector to register,
# produced by the pyg4ometry geometry-building code (omitted here)

# mark the detector physical volume as a germanium detector for remage
det_pv.set_pygeom_active_detector(
    RemageDetectorInfo("germanium", 101, {"manufacturer": "Mirion"})
)
# ...and change its appearance
det_lv.pygeom_color_rgba = (0.6, 0.6, 0.6, 1)  # for visualization

# define a custom region that contains the detector volume and set a
# custom production cut for it
region = Region("det_region")
region.add_volume(det_lv)
region.set_cuts(0.2 * u.mm)  # production cuts
region.add_to_gdml(reg)

# inspect the detectors previously defined in the Python code
for pv, det in walk_detectors(reg):
    print(pv.name, det.detector_type, det.uid)

# commit everything to a GDML file internally. This runs sanity checks to
# catch common geometry mistakes
write_pygeom(reg, "geometry.gdml")
\end{lstlisting}

\begin{lstlisting}[style=g4macro, caption={Example \remage macro that illustrates a typical simulation setup with registered detectors, vertex confinement and the selection of event kinematics. The geometry file is assumed to be set as a command-line option and is thus not spelled out in this listing.}, label={lst:remage-macro-example}]
# register a single physical volume as a Germanium-type detector with identifier 101
/RMG/Geometry/RegisterDetector Germanium hpge_01 101
# register all matched volumes individually, with id starting from 200
/RMG/Geometry/RegisterDetector Optical sipm_.* 200
# register all matching volumes into one single detector (with id 300)
/RMG/Geometry/RegisterDetector Calorimeter calo_segment_.* 300 .* true

# discard event if energy in detectors is < 100 keV
/RMG/Output/Germanium/EdepCutLow 100 keV
/RMG/Output/Germanium/AddDetectorForEdepThreshold 101 # select detector(s) by id

/run/initialize

# confine event vertices in a cylindrical sub-region of the "liquid_argon" volume
/RMG/Generator/Confine Volume
/RMG/Generator/Confinement/Physical/AddVolume liquid_argon
/RMG/Generator/Confinement/SamplingMode IntersectPhysicalWithGeometrical
/RMG/Generator/Confinement/Geometrical/AddSolid Cylinder
/RMG/Generator/Confinement/Geometrical/CenterPositionZ 0 m
/RMG/Generator/Confinement/Geometrical/Cylinder/InnerRadius 0 m
/RMG/Generator/Confinement/Geometrical/Cylinder/OuterRadius 3 cm
/RMG/Generator/Confinement/Geometrical/Cylinder/Height 5 cm

# alternative 1: generate event kinematics with a built-in Geant4 generator
/RMG/Generator/Select GPS
/gps/particle e-
/gps/energy 2.039 MeV
/gps/ang/type iso

# alternative 2: remage generator
/RMG/Generator/Select BxDecay0
/RMG/Generator/BxDecay0/DoubleBetaDecay Ge76 2vbb

# run the simulation
/run/beamOn 1000
\end{lstlisting}

\end{document}